\documentclass[12pt, preprint]{aastex}
\usepackage{emulateapj5}
\usepackage{psfig}

\def\alwaysmath#1{\ifmmode{#1}\else{$#1$}\fi}
\def\msun{\alwaysmath{\,{M}_{\odot}}}

\lefthead{Kundu et al.}
\righthead{Tidal Debris in the GGSS}

\begin{document}
\title{Exploring Halo Substructure with Giant Stars III: First Results from the Grid Giant Star Survey and Discovery of a Possible Nearby Sagittarius Tidal Structure in Virgo}

\author{A. Kundu\altaffilmark{1},
S. R. Majewski\altaffilmark{2,3},
J. Rhee\altaffilmark{2},
H. J. Rocha-Pinto\altaffilmark{2},
A. A. Polak\altaffilmark{2},
C. L. Slesnick\altaffilmark{2},
W. E. Kunkel\altaffilmark{4},
K. V. Johnston\altaffilmark{5},
R. J. Patterson\altaffilmark{2},
D. Geisler\altaffilmark{6},
W. Gieren\altaffilmark{6},
J. Seguel\altaffilmark{6}, 
V. V. Smith\altaffilmark{7}, 
C. Palma\altaffilmark{8},
J. Arenas\altaffilmark{6},
J. D. Crane\altaffilmark{2},
C. B. Hummels\altaffilmark{2}
}

\altaffiltext{1}{Michigan State University, Dept.\ of Physics \& Astronomy, East Lansing, MI 48824 (akundu@pa.msu.edu)}

\altaffiltext{2}{Dept.\ of Astronomy, University of Virginia,
Charlottesville, VA, 22903 (srm4n, rhee, hjr8q, aap5u, cls8q,
ricky, jdc2k, cbh4r@virginia.edu)}

\altaffiltext{3}{Visiting Research Associate, The Observatories of the
Carnegie
Institution of Washington, 813 Santa Barbara Street, Pasadena, CA 91101;
David and Lucile Packard Foundation Fellow; Cottrell Scholar
of the Research Corporation.}

\altaffiltext{4}{Las Campanas Observatory, Casilla 601, La Serena, Chile
(kunkel@jeito.lco.cl)}

\altaffiltext{5}{Wesleyan Univ., Dept.\ of Astronomy, Middletown, CT 06459
(kvj@astro.wesleyan.edu)}

\altaffiltext{6}{Grupo de Astronomia, Universidad de Concepci\'{o}n, Casilla 160-C, Concepci\'{o}n, Chile (doug@kukita.cfm.udec.cl, wgieren@coma.cfm.udec.cl, jseguel@andromeda.cfm.udec.cl, jose@gemini.cfm.udec.cl  )}

\altaffiltext{7}{Dept.\ of Physics, University of Texas, El Paso, TX 79968-0515, verne@balmer.physics.utep.edu}

\altaffiltext{8}{Dept.\ of Astronomy \& Astrophysics, The Pennsylvania State University, 525 Davey Lab, University Park, PA 16802, cpalma@astro.psu.edu}

\begin{abstract}

  We describe first results of a spectroscopic probe of selected fields
  from the Grid Giant Star Survey.  Multifiber spectroscopy of several
  hundred stars in a strip of eleven  fields along $\delta \approx
  -17^{\circ}$, in the range $ 12 \lesssim \alpha \lesssim 17$ hours,
reveals a group of
  8 giants
that have kinematical characteristics differing from the
  main field population, but that as a group maintain coherent, smoothly
  varying distances and radial velocities with position across the
  fields.  Moreover, these stars have roughly the same abundance,
  according to their MgH+Mgb absorption line strengths.  Photometric parallaxes place
 these stars in a semi-loop structure, arcing in a contiguous distribution between 5.7 and 7.9 kpc from the
  Galactic center.  The spatial,
  kinematical, and abundance coherence of these stars
  suggests that they are part of a
diffuse stream of tidal debris, and one 
roughly consistent with 
a wrapped, leading tidal arm of the Sagittarius dwarf spheroidal galaxy.
\end{abstract}


\section{Introduction}

        There has been a long and somewhat checkered history of observations
attempting to isolate substructure in the Galactic halo
\citep[e.g.,][]{E79,SC87,C91,C93,MMH96}. While observations in the last
decade have
confirmed the existence of a few coherent groups of halo stars
\citep{C93,MMH96,H99}, more recent large-area CCD surveys are beginning to
reveal the ubiquity of these structures \citep{V01,DP01,N02}. Given the
central importance of these probable relics of satellite accretion to the
\citet{SZ} model of the formation of the Galactic stellar halo, their
implications for the formations of massive halos via the Cold Dark Matter hypothesis
\citep{N97}, and their usefulness for measuring the Galactic
potential \citep{jzsh99}, there is motivation to identify 
and catalogue Galactic halo
streams and moving groups.  We present in this paper first
results from the Grid Giant Star Survey (GGSS), in which we identify
a new, coherent moving group of Galactic halo stars.

The GGSS probes 1302 evenly spaced, $\sim 0.4 - 0.6$
deg$^{2}$ areas of the sky to (a) supply several
thousand bright ($V \lesssim 12.5$), sub-solar metallicity K giants for
the Astrometric Grid of the Space Interferometry Mission \citep[see][]{conf},
 and (b) explore problems related to the structure and
kinematics of the Milky Way.
The photometric technique for
discriminating late type giants from dwarf stars, which uses the
Washington $M,T_2$ and $DDO51$ filters, is described by \citet{paperI},
and is being used by several groups \citep{mor00,gei02,paperII} to
explore Galactic structure problems.  Further discussion of the
reliability of the photometric selection of giant stars is given in
\citet{rebuttal}.  It bears reiteration here that follow up high resolution
spectroscopy -R=50,000; 5000-5900A$^{\circ}$ coverage; S/N$\approx$25 observations undertaken to monitor the spectroscopic stability of GGSS candidates and search for possible companions-  of stars selected
by the GGSS as metal-poor giants in a similar magnitude range as the stars discussed
here has
revealed 100\% of candidates observed so far to be, in fact,
metal-poor giants (Verne Smith, private communication). While preliminary inspection of $\approx$300 candidates suggests that they are all giants, detailed cross correlation studies with model spectra of a randomly selected
sub-sample of 38 stars confirms that they are indeed so with $-1.0 \lesssim Log(g) \lesssim 3.1$ and $T_{eff} \lesssim 5500 K$ for every candidate.

 GGSS photometry for 
$\delta < +20^{\circ}$ fields was obtained on the Swope 1-m telescope at
Las Campanas Observatory, and spans 
$9 \lesssim V \lesssim 17$.  Pipeline reductions
convert the three-filter data to photometric abundances and
parallaxes for 
likely giants according to the 
prescription in \citet{paperI}.  A more detailed discussion will be
given elsewhere.

We describe here first results from a spectroscopic probe of the
GGSS giant star sample from among the earliest fields 
processed through the photometric pipeline.  Samples of 
candidate 
giant stars in 11 fields along $\delta\approx-17^{\circ}$ (J2000)
and $ 12^{h}< \alpha<17^{h}$ were
observed with the
Blanco 4-m telescope + Hydra multifiber system on UT
28-30 Mar 2000.  We limited ourselves to
$M < 15$ giants for this
initial probe of GGSS fields because of limited available observing
time, obtained as brief ($< 1$ hour) hour angle ``fillers" for other projects.
The data were obtained using
the Schmidt camera and 2$\times$4k SITe CCD in conjunction with the KPGLD
grating. When centered on the infrared
calcium triplet region, this setup delivers radial velocity precision
of $ \sim 5$ km s$^{-1}$ (verified by repeat observations of test
stars).

\section{A Coherent Feature in Phase Space}

Fig. 1 shows the distribution of heliocentric radial velocities, $v_{helio}$ for all
photometrically classified giant (top) and dwarf (bottom) stars 
having Hydra spectroscopy, as a function of Galactic longitude. For comparison
the dotted curve marks the
minimum (maximum)
$v_{helio}$ for $270^{\circ}<l<360^{\circ}$  ($0^{\circ}<l<90^{\circ}$) expected for stars
within 20 kpc - the distance limit of our observations - if they follow a constant
circular rotation curve with $v_{rot}=$200 km s$^{-1}$ 
[we adopt a solar motion of $(u,v,w)=(-9,11,6)$ km s$^{-1}$ about a local
standard of rest with $v_{rot}=$220 km s$^{-1}$].
The solid line plots
the 0 km s$^{-1}$ ``rotation" curve, i.e. the reflex of solar motion.
These illustrative curves do not include the effects of velocity
dispersion about the adopted $v_{rot}$; nevertheless, while
the zero velocity dispersion and constant [$v_{rot}(R_{GC})$] simplifications
increasingly break down toward $l=0^{\circ}$ (where the bulge contributes),
the curves, which should nominally bound the expected $v_{helio}$
range for the Milky Way's flattened, disk-like components,
do make apparent several aspects of our sample.

For example, as expected, the velocity distribution of the dwarfs is
more consistent with a colder disk population than is 
that for giant stars, which tend to have a larger fraction of members
outside the curves.  Among these outliers we identify a conspicuous group of
eight stars, marked with solid circles in Fig. 1,
that trace 
a sequence of $v_{helio}$ (and $R_{GC}$, see below)
with position in the sky, with a smoothness that suggests that they are
members of a population having a tight velocity coherence (the progressive
variation 
reflecting primarily changes in line-of-sight perspective).
Table 1 lists 
observed and derived properties of these stars.
If a disk population,
these stars can only be matched by circular rotation curves with  $v_{rot}$
in the range $\sim 250-400$  km s$^{-1}$ at the distances reported in Table 1. This 
far exceeds the $v_{rot}$ of the Galaxy at the $R_{GC}$'s of this group (2-8 kpc).
The combination of kinematical {\it coldness} and rotational velocities
exceeding $v_{circ}$, which is generally only attained
near the perigalacticon of a highly elliptical orbit for halo objects,
supports the notion that these stars are part of a coherent debris stream from
a Galactic satellite.

        Fig. 2 shows the $(M-T_2, M-DDO51)$ distribution of the giant
candidates in Table 1 along with iso-metallicity loci of giant stars
[see \citet{paperI} for details].
Within the photometric uncertainties,
it can be seen that the stars in our proposed stream also appear to
have similar metal abundances of roughly [Fe/H] $=-1.0$.

	Fig. 3 shows the location of the stream stars in the sky; 
note that the positional coherence here is 
because of our study of fields along constant declination.
However, when the photometric parallaxes,
 $r$, and ($l,b$) positions of candidate stream stars are
converted to Cartesian Galactic coordinates, the stars
follow a
linear or arcing sequence, where each star is roughly equidistant from the
Galactic center (Fig. 4).
By accounting
for other dimensions of phase space, we have managed to ``follow" this
well-ordered sequence of stars into the velocity range occupied by more
typical field stars for $l<305^{\circ}$;
it is to the very existence of such a group of
stars with a smooth progression in various dimensions of parameter
space that we wish to draw attention, as it would be unusual to find
at random a group of stars that are 
velocity outliers,  have similar $R_{GC}$,
{\it and} show a
coherent sequence of velocities in {\it independently measured}
multifiber survey fields.

	To test whether this group of velocity outliers could be halo or thick disk stars 
aligned by chance, we performed a series of Monte Carlo simulations. 
For each GGSS field with a stream candidate (Table 1) we simulated
50,000 stars by assigning a random velocity defined by the halo velocity ellipsoid of \citet{C00}, and distance based on our distribution of photometric parallaxes. 
The probability of finding a halo outlier with radial velocity less than the velocity of a 
particular stream star (plus 5 km s$^{-1}$ to further account for velocity uncertainties) varies from 0.27 to 0.37 for
each GGSS field. The probability of finding eight outliers in the specific fields
listed in Table 1 is $\approx$0.0001. If we assume 
a probability of finding similar outliers in the three other Table 1 fields is also $\approx$0.3, 
then the probability of finding a set of outliers in eight consecutive fields is 0.0006. 
Similar simulations using the thick disk velocity ellipsoid from \citet{C00} in conjunction 
with a constant circular velocity of 200 km s$^{-1}$ yield probabilities of finding a random 
outlier from $<$0.001 to 0.35 in each of our fields, with a formal probability 
of a pure disk population indistinguishable from 0. Choosing the larger of the halo or thick disk outlier probability in each field yields a probability of at most 0.001 for finding outliers in eight consecutive fields.  We have an additional constraint in that the distance and 
velocity of our outliers are correlated (Table 1), with a correlation coefficient of 0.84. 
Assuming each of the eight stream fields has a star with the velocity listed in Table 1, we 
randomly assigned each star a distance from the ensemble of measured distances in each field. 
One hundred thousand simulations revealed only 2\% with an absolute value of the correlation 
coefficient greater than 0.84. Thus, we conclude that the probability of randomly finding a 
correlated stream of outliers, such as detected here, is negligibly
small (probability of 0.001$\times$0.02). Thus the combined evidence of a smooth, cold progression in
the various projections of phase space ($l-v_{helio}$, $l-b$, $X-Z$, $b-r$,
$r-v_{GSR}$ -- see Figs. 1, 3 and 4) strongly suggests that
these stars are parts of a loop of tidal debris arcing around the Galactic
Center.

We looked for other such ``sequences" in the data but found nothing else as striking.   
There are other velocity outliers in Fig. 1, as would be expected
from our estimate of a $\approx$30\% chance to find a velocity outlier in any particular field; 
but, for example, the four positive velocity outliers in
Fig. 1 show no obvious correlation with distance, and yield a metallicity dispersion twice that
of our proposed stream. 

        Our observations suggest the Table 1 stream to be extremely diffuse. Each
GGSS pointing used for this set of observations covers an area of 0.4 deg$^2$
separated by $5.6^{\circ}$ from its neighbor. The absolute magnitudes of both our stream 
and non-stream giants span the entire range covered by the ``red giant branch bump"  and 
``normalization region" stars reported by Bono et al. (2001, ``B01" hereafter) in their 
HST based starcount studies of Galactic globular clusters.
By comparing the stellar density of our stream stars with the total density of ``red giant branch bump" 
and ``normalization region" stars in the B01 globular clusters we estimate
the difference/offset in surface brightness densities.  The globular cluster NGC 6362 
has an average metallicity in the range of our proposed stream stars,
and is a diffuse, nearby object \citep{H96} -- so that the average density/surface
brightness within the HST-WFPC2 field of view of the B01 observations
is roughly constant, and approximately equal to the central value. By adding the
difference in surface brightness density of the stream with respect to NGC 6362, calculated 
above, to the central surface brightness of this globular cluster \citep{H96} 
 we estimate the surface brightness of the stream to be
fainter than $\sim 29$ mags arcsec$^{-2}$ in the $V$ band.
While this estimate is admittedly crude, it is likely that our stream is among the
most diffuse
known coherent substructures in the Galactic halo, and this find highlights
the capability of the GGSS to discern
such a gossamer thread among the clutter of Galactic stars.
We note that our ``tidal feature" does not seem to correspond
well with any of the features in the Sloan Digital Sky Survey
(SDSS) data of \citet{N02},
even though their observational plane is only 17$^{\circ}$
different from ours.  This may be attributable both to the low surface
brightness of the feature,
as well as the fact that the main sequence turn-off counterparts
of the stars we see here are at the bright magnitude limit of the
SDSS.

        The extent of the proposed debris stream is presently not well defined,
but we are undertaking follow-up photometry and spectroscopy to
bracket the limits better.  Note that we lose track of the
sequence past $l=345^{\circ}$ possibly due to magnitude limitation -- the $l=350^{\circ}$
field would have the next farthest star from us, but this is close to the
magnitude limit of the present spectroscopic sample.   At the other end of the
sequence, any progression of the sequence of stars is lost among
disk stars having similar velocities.
It is interesting that if our sequence is indeed tidal debris,
then extrapolation brings it
through the solar neighborhood.

\section{Possible Association with Sagittarius }

The positions of our selected stars in Fig. 3 and the top left panel of
Fig. 4  are very suggestive of tidal
debris from a satellite orbiting the Galaxy on a nearly polar orbit.  Of
the galaxies known to be satellites of the Milky Way today, the
Sagittarius dwarf \citep[discovered by][]{igi94} is the one most
likely to be the parent of this debris since it is the closest of all
the satellites to the Galactic center (currently at 16 kpc) and is also
on a nearly polar orbit passing through the region where our fields
are located. Not only is the highly distorted shape of the main
body of this galaxy suggestive of tidal disruption \citep{i+97}, but
there is already substantial evidence for significant debris associated
with this galaxy at many tens of degrees from its main body
\citep{mom98, starcountsIII, iils01, ive00, dmgc00, sdss00, magc01,V01,DP01,N02}.
We note that the Table 1 stars have photometric [Fe/H] that approximate the
-1.3 and -0.7 dex values of the 11 and 5 Gyr aged populations seen in
Sgr \citep{LS00}.

Figs. 3 and 4 compare our data with semi-analytic predictions for the
position and velocity of debris lost from Sgr during the last 1.5 Gyrs
\citep[see][for a description of the model]{KVJ98}. In this particular
model Sgr had an initial mass of 2.1$\times$10$^{9}$ $\msun$, the Milky Way's circular velocity was fixed
at 200 km s${}^{-1}$, and Sgr's tangential velocity was allowed to vary
within the proper motion error bars of present day observations to find the best fit, $u=200$, $v=-20$ and $w=175$ km s$^{-1}$.   The portion
of the streamer that is the closest match to the data comes from
debris lost $\approx$1 Gyr ago, two pericentric passages previous to the most recent encounter i.e. we adopt perigalacticon passage n=0 for the most recent encounter  while the passage of interest in n=-2.  The debris from this encounter has wrapped entirely around the Milky Way.

 Figs. 3 and 4 suggest that it is plausible that the feature we have found
is indeed associated with Sagittarius. However, we note that this feature
alone cannot strongly constrain the details of Sgr's orbit, mass, mass
loss history or Milky Way potential. The profusion of streams plotted in
our figures is some indication of the degeneracy of the problem.
Rather, our discovery is interesting in that it is the first case
reported of debris that could be from Sagittarius within the Sun's orbit
and, if the association is correct, suggests that there should be many
more Sgr streams wrapped around the Galaxy within a few kiloparsecs of the Solar
Neighborhood. Previous studies 
 \citep{TI98,dmgc00,J99} 
have
suggested evidence of multiply wrapped tails; the GGSS can provide a direct
means of identifying and studying such features in the inner halo.
Considering that our first detailed look at a portion of the GGSS data
revealed evidence of such a diffuse stream we are hopeful that future
studies with the GGSS data set will yield more such features, as part
of our goal to understand the formation and mass distribution 
of the Galactic halo.

Note added in proof: A preliminary reduction of new follow up 
spectroscopic  
observations (of a mix of stars previously observed, and 
fainter giant candidates that we hadn't observed spectroscopically 
earlier) obtained since the paper was submitted shows more 
outliers consistent with the stream. They even extend the stream 
out to the fields G1532-16 and G1556-16 and have stream members 
with even more extreme outlying velocities.

\acknowledgements

Our work on the Grid Giant Stars Survey has been supported by NASA/JPL
by the following SIM Preparatory Science Grants:
Grant \#AST-1201670 (SRM as PI), Grant \#AST-1222563
(SRM as PI), and Grant \#NRA-99-04-OSS-058 (DG as PI).  SRM also
acknowledges support from NSF Career Award AST 97-02521, a Cottrell Scholar
Award from The Research Corporation, and a David and Lucile Packard Foundation
Fellowship.

\begin{figure}[!hb]
\plotone{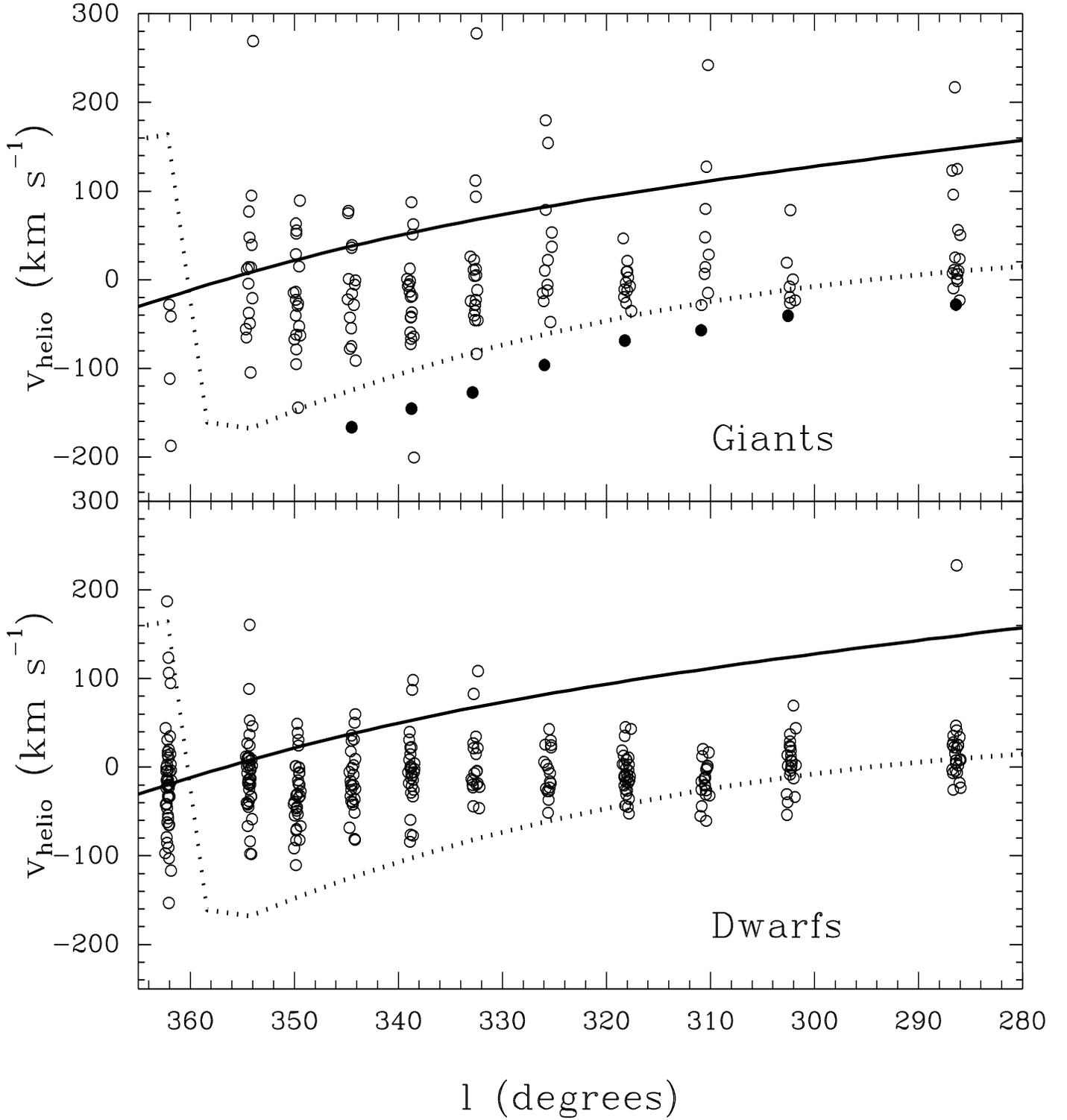}
\caption{Top: Galactic longitude vs. $v_{helio}$ for
candidate giants. Bottom: Corresponding plot for dwarfs.
The dotted line plots the largest $v_{helio}$ 
expected for a population of stars within $r = 20$ kpc
uniformly rotating at $v_{rot}$ = 200 km s$^{-1}$ for all $R_{GC}$.
The solid line traces the 0 km s$^{-1}$ ``rotation" curve.
Solid dots identify the candidate stream. 
}
\end{figure}

\begin{figure}[!hb]
\plotone{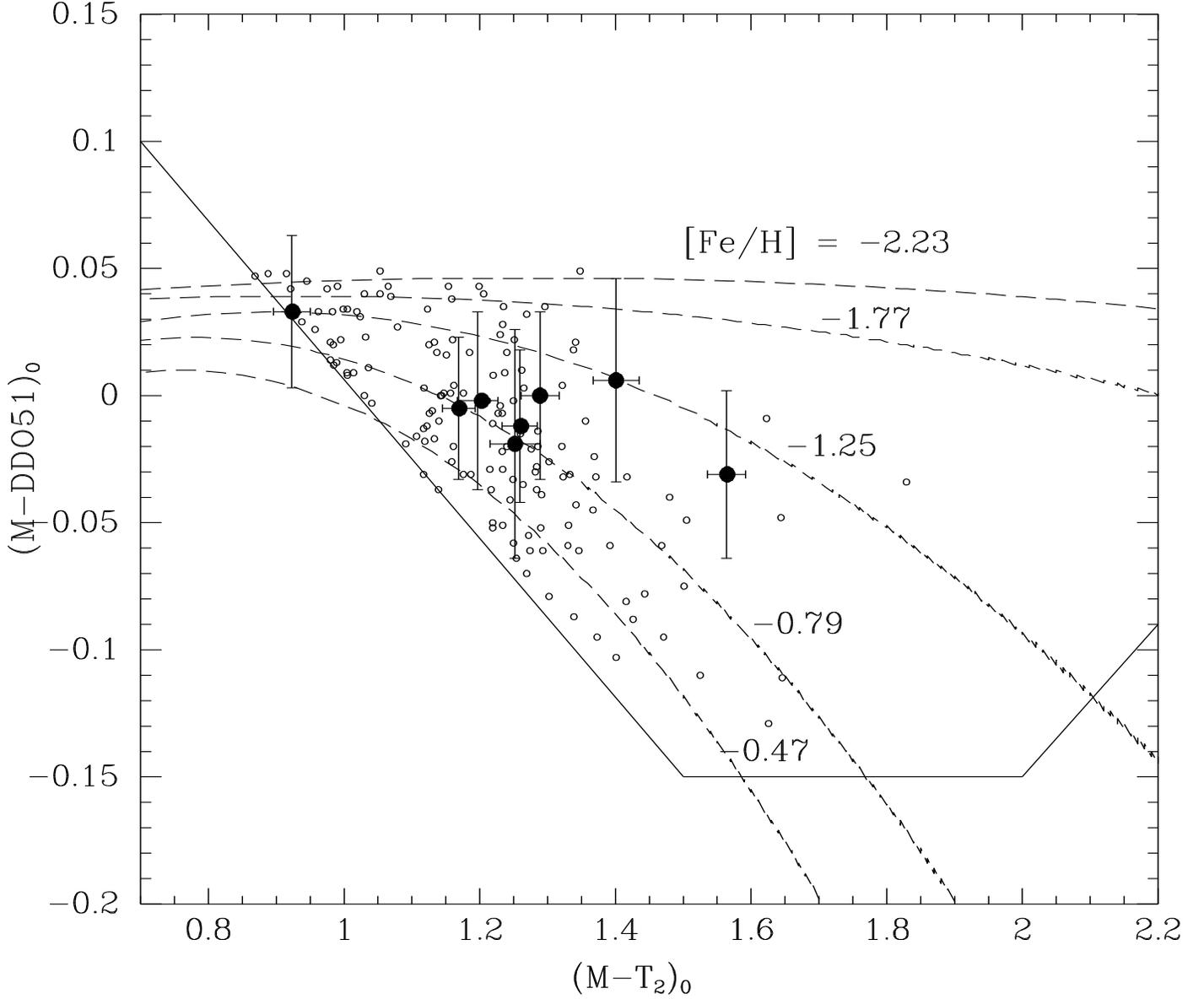}
\caption{$(M-T_2, M-DDO51)$ distribution of photometrically identified giants.  
The iso-metallicity loci are 
from Majewski et al.\ (2000a).
All candidate stream giants (solid circles) share
similar metallicities in a small range around [Fe/H] = -1.0.}

\end{figure}

\begin{figure}[!hb]
\plotone{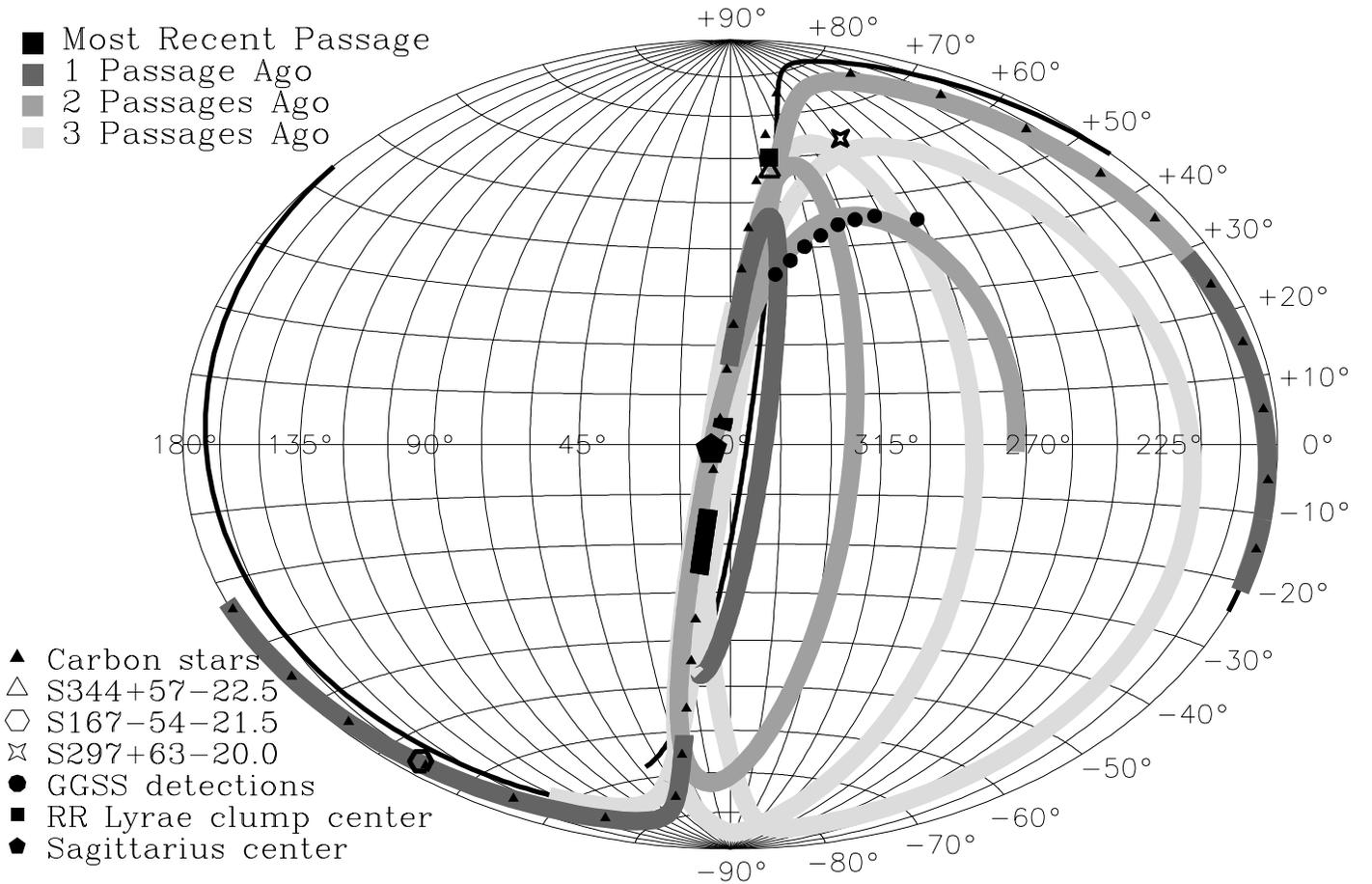}
\caption{The positions of Sgr debris tails in ($l,b$) space from the
Johnston (1998) semi-analytical model that best fits previous observations of
tidal Sgr debris (see Johnston et al. 1999). The thin solid line traces the past motion of the Sgr dwarf. 
The proposed stream is most consistent with the debris
stripped from Sgr ''two passages ago". }
\end{figure}

\begin{figure}[!hb]
\plotone{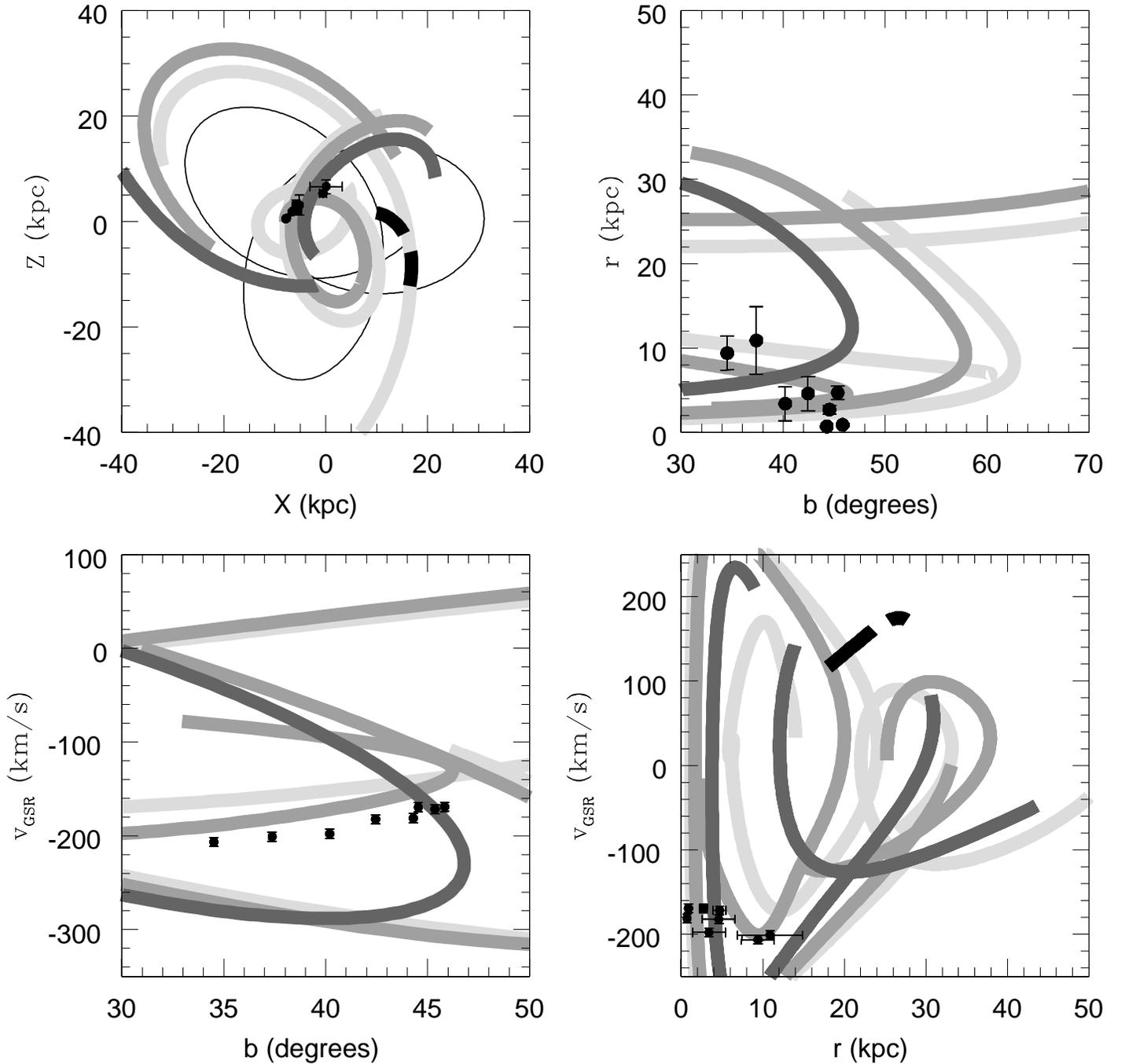}
\caption{ The tidal streams predicted by semi-analytical models, along with the
data points for the proposed stream
 in various coordinate combinations. Lines as in Fig. 3. For clarity, the past orbit of Sgr is only shown in upper left plot.
Top left: Galactic Cartesian coordinates $X$ vs. $Z$, with the Galactic center at (0,0).
Top right: $b$ vs. heliocentric distance.
Bottom left: $b$ vs.  radial velocity corrected to the GSR system.
Bottom right: Heliocentric distance vs. $v_{GSR}$. 
Only streams in the range $280^{\circ}<l<370^{\circ}$ are shown for the $b$-$r$,  and $b$-$v_{GSR}$ 
plots in order to reduce confusion. }
\end{figure}

\begin{deluxetable}{llllllllll}
\tabletypesize{\scriptsize}
\tablewidth{0pt}

\tablenum{1}
\tablecaption{Salient Properties of Candidate Stream Giants\label{table1}}
\tablehead{Name    & l,b\tablenotemark{a} & $\alpha$\tablenotemark{b} & $\delta$\tablenotemark{b}    & $M$ 	    &   $(M-T)_o$   & $(M-D)_o$ 		 & $v_{helio}$ 	 & [Fe/H] 	  & Distance \\
        &  &   & &\colhead{(mag)}   &  &  & \colhead{(km s$^{-1}$)} &  & \colhead{(kpc)}    }
\startdata
 G1200$-$16.68 & 286.5,44.2	& 12:02:16.84    & $-$17:00:08.7   & 11.70$\pm$0.02 & 1.17$\pm$0.02 & $-$0.01$\pm$0.03 		 &   \phantom{1}$-$28.2  & $-$0.8$\pm$0.4 &    \phantom{1}0.7$\pm$0.5 \\
 G1247$-$16.160 & 302.4,45.7	& 12:50:18.88   & $-$17:02:40.3   & 13.48$\pm$0.02 & 0.92$\pm$0.03 & \phantom{$-$}0.03$\pm$0.03 &   \phantom{1}$-$40.7  & $-$1.3$\pm$2   &    \phantom{1}0.9$\pm$0.5 \\
 G1312$-$16.107 & 310.5,45.4	& 13:14:51.53   & $-$17:09:20.0   & 13.97$\pm$0.02 & 1.29$\pm$0.03 & \phantom{$-$}0.00$\pm$0.03 &    \phantom{1}$-$57.0 & $-$1.1$\pm$0.4 &    \phantom{1}4.7$\pm$2 \\
 G1334$-$16.68 & 318.2,44.4	& 13:36:46.14    & $-$16:59:05.6   & 13.43$\pm$0.02 & 1.26$\pm$0.03 & $-$0.01$\pm$0.03 		 &   \phantom{1}$-$68.8  & $-$0.9$\pm$0.4 &    \phantom{1}2.7$\pm$1 \\
 G1359$-$16.66 & 325.7,42.7	& 14:01:49.01    & $-$17:15:25.0   & 15.03$\pm$0.03 & 1.25$\pm$0.04 & $-$0.02$\pm$0.04 		 &   \phantom{1}$-$96.3  & $-$0.8$\pm$0.5 &    \phantom{1}4.6$\pm$3 \\
 G1423$-$16.112 & 332.6,40.4	& 14:25:28.89   & $-$17:08:20.5   & 14.42$\pm$0.02 & 1.20$\pm$0.03 & $-$0.00$\pm$0.04 		 &  $-$127.3  		 & $-$0.9$\pm$0.5 &    \phantom{1}3.3$\pm$2 \\
 G1445$-$16.181 & 338.9,37.5	& 14:47:59.37   & $-$17:16:16.8   & 14.91$\pm$0.03 & 1.40$\pm$0.03 & \phantom{$-$}0.01$\pm$0.04 &  $-$145.6  		 & $-$1.2$\pm$0.4 &   10.9$\pm$4 \\
 G1507$-$16.326 & 344.6,34.2	& 15:10:45.88   & $-$16:53:19.5   & 14.09$\pm$0.02 & 1.56$\pm$0.03 & $-$0.03$\pm$0.03 		 &  $-$166.5  		 & $-$1.1$\pm$0.2 &    \phantom{1}9.0$\pm$2 \\
 G1532$-$16\tablenotemark{c}     & 349.7,30.6	& \multicolumn{1}{c}{\nodata}  & \multicolumn{1}{c}{\nodata} &\multicolumn{1}{c}{\nodata} &\multicolumn{1}{c}{\nodata} &\multicolumn{1}{c}{\nodata} &\multicolumn{1}{c}{\nodata} &\multicolumn{1}{c}{\nodata} &\multicolumn{1}{c}{\nodata}\\
 G1556$-$16\tablenotemark{c}     & 354.2,26.6	& \multicolumn{1}{c}{\nodata}  & \multicolumn{1}{c}{\nodata} &\multicolumn{1}{c}{\nodata} &\multicolumn{1}{c}{\nodata} &\multicolumn{1}{c}{\nodata} &\multicolumn{1}{c}{\nodata} &\multicolumn{1}{c}{\nodata} &\multicolumn{1}{c}{\nodata}\\
 G1643$-$16\tablenotemark{c}     & 2.2,18.0	& \multicolumn{1}{c}{\nodata}  & \multicolumn{1}{c}{\nodata} &\multicolumn{1}{c}{\nodata} &\multicolumn{1}{c}{\nodata} &\multicolumn{1}{c}{\nodata} &\multicolumn{1}{c}{\nodata} &\multicolumn{1}{c}{\nodata} &\multicolumn{1}{c}{\nodata}\\
 \enddata
\tablenotetext{a} {Field centers of observations}
\tablenotetext{b} {Position of candidate stream giants in J2000 coordinates.}
\tablenotetext{c} {No candidate stream stars were found in these fields}
\end{deluxetable}

\end{document}